\documentclass[aps,prl,twocolumn,superscriptaddress,groupedaddress,showpacs]{revtex4-1}  
\usepackage{relsize}
\usepackage{cancel}
\usepackage{mathtools}
\usepackage[T1]{fontenc}
\usepackage{geometry}
\usepackage{fancyhdr}
\usepackage{hyperref}

\usepackage{bibentry}

\usepackage{color}

\usepackage{graphicx}
\usepackage{dcolumn}
\usepackage{bm}
\usepackage{enumerate}
\usepackage{slashed}
\usepackage{simplewick}
\usepackage{comment}

\usepackage{soul}

\usepackage[english]{babel}
\usepackage{appendix}
\usepackage[utf8]{inputenc}

\usepackage{marginnote}
\usepackage[normalem]{ulem}

\usepackage{subfigure}

\newcommand*\xbar[1]{%
   \hbox{%
     \vbox{%
       \hrule height 0.5pt 
       \kern 1.0ex
       \hbox{%
         \kern-0.1em
         \ensuremath{#1}%
         \kern-0.1em
       }%
     }%
   }%
} 
\newcommand*\xxbar[1]{%
   \hbox{%
     \vbox{%
       \hrule height 0.5pt 
       \kern 0.8 ex
       \hbox{%
         \kern-0.1em
         \ensuremath{#1}%
         \kern-0.1em
       }%
     }%
   }%
}

\newcommand{\eq}[1]{(\ref{#1})}

\newcommand{\la}{\label}
\newcommand{\ba}{\begin{align}}
\newcommand{\ee}{\end{equation}}
\newcommand{\be}{\begin{equation}}

\def\12{\frac{1}{2}}
\newcommand{\p}{\partial}
\newcommand{\en}{\end{align}}
\newcommand{\e}{\epsilon}

\usepackage{color}

\usepackage{amsmath}

\begin{document}

\setcounter{secnumdepth}{-1} 

\title{Axial-Current Anomaly in Euler Fluid   }

\author{A.G.~Abanov}

\affiliation{Department of Physics and Astronomy and Simons Center for Geometry and Physics, Stony Brook University, Stony Brook, NY 11794, USA}
\author{P.B.~Wiegmann}
\affiliation{Kadanoff Center for Theoretical Physics, University of Chicago, 5640 South Ellis Ave, Chicago, IL 60637, USA}
\affiliation{Institute for Information Transmission Problems, Moscow, Russia}

\date{\today}

\begin{abstract}
We argue that a close analog of the axial-current anomaly of quantum field theories with fermions occurs in the classical Euler fluid. The conservation of the axial current (closely related to the helicity of inviscid  barotropic flow) is anomalously broken by the external electromagnetic field as $\p_\mu j_{A}^\mu = 2\,\bm E\!\cdot\! \bm B$ similar to that of the axial current of a quantum field theory with Dirac fermions such as QED. 

\end{abstract}

\date{\today}

 \maketitle

\newpage

\paragraph{Introduction.} 
 Axial-current anomaly of QED asserts
that while the electric (vector) current
of Dirac fermions \(j^\mu=\bar\psi\gamma^\mu\psi\) is conserved, the axial
current \(j^\mu_A=\bar\psi\gamma_5\gamma^\mu\psi\) is not
\begin{align}
        &\p_\mu j^\mu=0\,,
 \la{01}\\
        &\p_\mu j_A^\mu=\tfrac k4 {}^\star\! F F\,,
 \la{1}
\end{align}
where \({}^\star\! F^{\mu\nu}=\tfrac
12\epsilon^{\mu\nu\lambda\rho} F_{\lambda\rho}\) 
is the dual electromagnetic tensor. The constant \(k\) is integer valued when electromagnetic tensor  \(F\) is measured in units of the magnetic flux quantum \(\Phi_0=hc/e\). In QED
it is \(k=2\), the number of Weyl fermions in the Dirac multiplet. In
terms of electric and magnetic field the anomaly \eq{1} reads
\be  \p_\mu j_A^\mu=k\bm E\!\cdot\! \bm B \,.
\la{00}\ee

The term {\it anomaly} emphasizes that while simultaneous  transformation
\(\psi_{L,R} \to  e^{i\alpha}\psi_{L,R}\) of left and right components of
the Dirac multiplet
by virtue of the Noether theorem yields the conservation of  the electric
charge \(Q=\int j^0d \bm x\), the axial transformation
\be
        \psi_{L,R}\to e^{\pm i\alpha_A}\psi_{L,R} 
 \la{002}
\ee
does not warrant the conservation of the chirality \(Q_A=\int j^0_Ad \bm
x \) even though it leaves the classical Dirac equation unchanged. It follows
from \eq{1} that  
\begin{align}
        \tfrac d{dt}Q=0\,,\quad\tfrac d{dt}Q_A=2\int\bm E\!\cdot\!\bm Bd\bm x \,. 
 \la{02}
\end{align}
Obtained in 1969 by Adler \cite{Adler} in QED and Bell and Jackiw \cite{Bell-Jackiw} in the liner \(\sigma\)-model the axial-current anomaly (or PCAC, partial conservation of axial current) is a fundamental non-perturbative result in gauge field theories which goes well beyond QED, proven experimentally at different scales of high energy. The most recent advances are taken place in heavy-ion collision  \cite{Kharzeev}, the field  which initiated a search for anomalies in relativistic hydrodynamics \cite{Son} (see, also \cite{Abanov}). Anomalies also  have important applications in quantum  fluids, the most notably in the superfluid $He^3$ \cite{Volovik}.

\paragraph{Main Results.} In this paper
we show that axial-current anomaly is also a property of a \underline{classical
Euler's hydrodynamics} of the ordinary inviscid barotropic  fluid.
Such
fluid is described by the Euler equations with  Lorentz force
\begin{align}
        \!\!&\dot\rho+\bm\nabla(\rho\bm v)=0 \,,
 \la{003} \\ 
        &(\p_t+\bm v\!\cdot\!\bm\nabla)m\bm v+\bm\nabla \mu
        =e\bm E+(e/c)\bm v\times \bm B\,,
 \la{03}
\end{align}
 where \(\mu\), a function of the density \(\rho\), is the chemical potential related to pressure as $dp=\rho d\mu$. The fluid assumed to be electrically charged responding  to electromagnetic field.

Like QED, the barotropic fluid possesses two locally conserved charges. One is electric charge (the mass in units of \(e\)) \(Q=\int \rho d\bm x\). Its current, a 4-vector \(j^\mu=(\rho,\ \rho\bm v)\) is manifestly divergence-free as it is  stated by the continuity equation \eq{003} and expressed by \eq{01}.

The axial charge is the fluid helicity defined in  \cite{Moffat}
as
\begin{align}
 \mathcal{H}=(1/\Gamma^{2})\int 
 \bm v\cdot (\nabla\!\times\! \bm v)d 
 \bm x \,.
 \la{05} 
\end{align}
It is conserved in the absence of external fields. If we assume that the vorticity is concentrated in thin vortex (closed) filaments of an equal circulation $\Gamma$ the helicity is twice the linking  number of the filaments \cite{Moffat}. In superfluid \(\Gamma=h/m\)  is Onsager circulation quantum ($h$ is  the Planck constant). Even though we deal with classical flows, we normalize the  helicity \eq{05} by the Onsager circulation quantum. This brings the normalization close to that of fermionic quantum theories.

Furthermore, we adopt the units where \(h,e,c=1\), but keep \(m\) to distinguish between the fluid momentum and the velocity. That distinction has a profound physical significance in the presence of external fields when the difference between momentum and velocity becomes important. In this case,  the helicity is defined through the density of canonical momentum 
\be 
    \bm\pi = m\bm v +\bm A
 \la{pidef}
\ee  
as 
\be \mathcal{H}=\frac 1{h^2} \int\bm
        \pi\cdot(\bm\nabla\times
        \bm \pi) d\bm x \,,
 \la{3}
\ee
where \(\bm A\) is the electromagnetic vector potential. 

We comment that while the fluid momentum (\ref{pidef}) is defined up to a gradient of a function, the helicity (\ref{3}) is uniquely  defined  for appropriate boundary conditions (e.g., a closed manifold).

With or without external field, the helicity defined by \eq{3} is conserved
\be
        \tfrac d{d t} {\mathcal{H}}=0\,.
 \la{06}
\ee
However, the helicity density $\bm h_0=\pi\cdot(\bm\nabla\times \bm \pi)$ depends locally on the gauge potential and cannot be treated as a local Eulerian field.

We argue that in hydrodynamics the axial charge and its density should be identified with 
\begin{align}
    Q_A &=\int\rho_Ad \bm x \,,
 \la{6} \\
    \rho_A &=m\bm v\!\cdot\!\left(\bm \omega +2\bm B\right)\,,\;\;\; \bm \omega=\bm\nabla\!\times(m\bm v)\,,
\la{9}
\end{align}
where $\bm \omega$ (defined as a curl of the fluid momentum) is the vorticity of the fluid. We refer to $Q_A$ as {\it a fluid chirality}  bringing the terminology closer to that of QED.   Chirality is the sum of the fluid helicity (the first term in (\ref{9})) and twice the cross-helicity (the second term). In contrast to the helicity density, the chirality density is a local Eulerian field.
  
We will show that the chirality (\ref{6}) obeys the anomaly equation \eq{02}, while the chirality density \eq{9} obeys the local anomaly equation \eq{1}
\begin{align}
  \dot\rho_A+\bm\nabla \bm j_A=2 \bm E\!\cdot\!\bm B \,,
 \la{09}
\end{align} 
where the chirality flux \( \bm j_A\) (defined modulo a curl) is explicitly given by 
\begin{align}
        \!\!\!\!\!\bm j_A\!=\!\rho_A\bm v\!+\!(\bm\omega\!+\!2\bm {B})
        \left(\!\mu\!-\!\frac{m\bm v^2}{2}\!\right)\!-\! m\bm v\!\times\!(\bm {E}\!+\!\bm v\!\times\!\bm {B}) .
 \la{8}
\end{align}

The chosen normalization identifies the factor 2 in \eq{9} and \eq{09} with the value of the triangle diagram in QED. The latter is the coefficient \(k=2\) in the anomaly equation \eq{1} which is 
 itself is a topological number \cite{Volovik}. Later we discuss the relationship between the anomaly and linking numbers and briefly touch on the topological interpretation of this factor (see (\ref{10}-\ref{11})). 

Eqs.~(\ref{9},\ref{8}) could be considered as a realization of the anomaly equations \eq{1} by hydrodynamics similar to the Wess-Zumino non-linear \(\sigma\)-model \cite{WZ}. Also, the `bosonization' of Dirac fermions in one spatial dimension could be seen as a precursor of an axial current anomaly in hydrodynamics in higher odd spatial dimensions as Eqs.~(\ref{9},\ref{8}) demonstrate. Similarly, the axial anomaly appears in non-inertial reference frames. For example, in the case of rotating fluid subject to a potential force, say, gravity, one replaces the magnetic field with the frequency of rotation in the equations above. 

There is a simple physical picture behind these formulas. Consider a fluid
in electric and magnetic fields in a local reference frame moving and rotating with the fluid. In this frame the chirality density \eq{9} is locally approximated by \(\rho_A=2m\bm v\!\cdot\!\bm B\) and the chirality flux \(\bm j_A\) is divergent free. Electric field accelerates the fluid \(m\dot{\bm v}=\bm E \). Hence \(\dot\rho_A=2(\bm E\!\cdot\!\bm B)\) as in \eq{09}. Going back to the laboratory frame the magnetic field transforms \(\bm B\to\bm B+\bm\omega/2\) (Larmor precession) and the formula for chirality transforms as \(\rho_A=2m\bm v\!\cdot\!\bm B\to m\bm v\cdot(\bm\omega+2\bm B)\). At the same time, \(2( \bm E\!\cdot\!\bm B)\) being an invariant undergoes no change. This yields the formulas \eq{9} and \eq{09}. The term \(2\bm B\mu\) in \eq{8} and the extension \(2\bm B\mu\to (\bm\omega+2\bm B)\mu\)  is reminiscent {\it the chiral magnetic effect} \cite{Kharzeev} and  {\it the chiral vortical effect} of Ref.~\cite{kharzeev2011testing}, although in these  papers the term \((\bm\omega+2\bm B)\) appeared in the vector current, not in the axial current as in \eq{8}.

To elucidate the global aspect of the axial anomaly in hydrodynamics we assume that the vorticity and magnetic field are approximated by vortex and magnetic filaments having the same circulation \(\Gamma\) and the same magnetic flux \(\Phi_0\), respectively, as if the fluid were a superfluid (our calculations do not rely on the assumptions of discreteness but are more of a kinematic nature). The Helmholtz law warrants that once created, vortex lines and their bundles can not be destroyed.
 If vortices and magnetic lines are discrete then in units \(\Gamma\) and \(\Phi_0\) the fluid helicity \(\int (\bm v\cdot\bm\omega)d \bm x\) \eq{3}, the magnetic helicity \(\int(\bm A\cdot\bm B)d\bm x\), the cross-helicity \(\int(\bm v\cdot\bm B)d \bm x\) are topological invariants. In the respective order they are twice the linkages between vortex lines \({\rm 2\,Link}[\bm\omega]\) \cite{Moffat}, magnetic flux lines \({\rm 2\,Link}[\bm B]\) \cite{MoffattRicca} and the cross-helicity is the mutual linkage between vortex and magnetic lines \({\rm Link}[\bm\omega,\bm B]\) \cite{Moffatt78}. Hence, the chirality \(Q_A\) \eq{6} and helicity \eq{3} are even integers written as sums of the linkages
\begin{align}
  Q_A &=2{\rm \,Link}[\bm\omega]+2{\rm \,Link}[\bm\omega,\bm B]\,,
 \la{10} \\
  \mathcal{H}\ &=2{\rm \,Link}[\bm\omega]+2{\rm \,Link}[\bm\omega,\bm B]+{\rm 2\,Link}[\bm B]\,.  
 \la{13}
\end{align} 
Then, the relation between chirality and helicity is
\begin{align}
  &Q_A= \mathcal{H}-2{\rm \,Link}[\bm B] \,. 
 \la{11}
\end{align}
While the individual linkages in the RHS of \eq{13} may change in the course of the flow, the total helicity $\mathcal{H}$ does not (see (\ref{06})). Then the time derivative of
(\ref{11}) gives the anomaly equation \eq{02}.
If we treat the chirality \(Q_A\) as a difference between the number of left and right moving fermions as in QED, we may say that an extra link to magnetic lines in accord to \eq{11} changes the chirality by 2, by flipping the chirality of a fermion from right to left, and according to \eq{13} changes the sum of linkages of the vortex lines and the mutual linkage by 1 \cite{Note1}.

 As an example of the linkage changing evolution, consider an instantaneous process of formation of a closed magnetic filament with magnetic flux \(\Phi_0\). A fast  change of magnetic field triggers a strong electric field that spins the fluid around the filament. As a result, the vorticity loop is formed along the magnetic filament. The magnetic field and vorticity thus created satisfy $\bm\nabla\times \bm \pi =\bm\omega+\bm B=0$. The magnetic and vortical linkages are changed in that process while the total helicity (\ref{3},\ref{13}) does not. 

Eqs.(\ref{9}-\ref{8}) and their global version (\ref{10}) are the major results of this study.

Since Eqs.(\ref{9}-\ref{8}) follow from the Euler equation
\eq{03} one can validate them by elementary means being equipped by no more than the vector calculus. Below are the evolution equations for helicity  and cross-helicity densities obtained directly from the Euler equations \eq{03}
\begin{widetext}
\begin{align}
        \begin{split}
        &\p_t(m\bm v\cdot\bm {B})+\bm\nabla\left[\bm v(m\bm v\cdot\bm {B})
        +\bm {B}\left(\mu-\frac{m\bm v^2}{2}\right)-m\bm v\times(\bm {E}+\bm v\times\bm {B})\right] 
        +\bm\omega\cdot(\bm{E}+\bm v\times\bm {B}) = \bm {E}\cdot \bm {B}\,,
 \\
 	& \p_t(m \bm v\cdot \bm\omega) 
        + \bm\nabla\left[\bm v(m\bm v\cdot\bm\omega)
        +\bm\omega\left(\mu-\frac{m\bm v^2}{2}\right)
        +m\bm v\times(\bm {E}+\bm v\times\bm {B})\right] 
        -2\bm\omega\cdot(\bm {E}+\bm v\times\bm {B}) =0\,.
        \end{split}
 \la{17}
 \end{align}
\end{widetext}
Combining we obtain the equation identical to (\ref{9}-\ref{8}) with the
helicity flux of the form equivalent to \eq{8}.

We emphasize that we do not discuss magnetohydrodynamics (MHD) in this work. The formulas \eq{06} and \eq{09} have been derived when the electromagnetic field is treated as an external having no feedback from the charged fluid motion.  The setting similar to the one we consider occurs in the regime  referred as Hall MHD, when the Lorentz force in Eq. \eq{03} acting on fluid of ions is controlled by the fast  motion of electrons largely independent from the ions flow \cite{Lighthill}. Contrary, in the limit of ideal (i.e., infinitely conducting)  MHD the Ohm's law yields $\bm {E}+\bm v\times\bm {B}=0$ and consequently \(\bm E\!\cdot\! \bm B=0\). In this case, the cross-helicity and helicity densities conserve separately as it follows from (\ref{17}). In plasmas with finite conductivity, only the total helicity is conserved.

Behind the straightforward algebra yielding to (\ref{17}) and (\ref{9},\ref{09},\ref{8}) there are deeper symmetry and geometry-based reasons (see \cite{Arnold} as a general reference for geometric view on hydrodynamics). Here we only touch the surface leaving a more comprehensive discussion to future publications. We start from the derivation of the helicity conservation in the form that makes the conservation of chirality an easy corollary.

\paragraph{Vorticity transport and helicity conservation.} We will use the 4-dimensional space-time formalism. The formalism is standard in relativistic hydrodynamics but is not common in studies of   non-relativistic  flows. Still, we find that it is the most compact way to expose the geometric nature of  fundamental laws of the Euler flow: Helmholtz law for advection of vorticity and the conservation of helicity with or without external fields and in non-relativistic or relativistic hydrodynamics alike. A reason is that these laws are expressed in terms of differential forms and, therefore, are not sensitive to the space-time metric.  

We start by writing  the Euler  equation \eq{03} in terms of the fluid momentum 
 \begin{align}
    \rho(\dot{\bm \pi}-\bm\nabla\pi_0)-\rho\bm v\times(\bm\nabla\times\bm \pi)=0 \,.
 \la{028}
\end{align} 
Here \(\pi_0\) denotes the Bernoulli function \(\)
\begin{align}
    \pi_0=\Phi+A_0\,,\quad -\Phi=\mu+\tfrac 12 mv^2 \,,
 \la{pi0def}
\end{align} 
where \(A_0\) is the electrostatic potential.

Next we recognize the mass 4-current \(j^{\mu} = (\rho,\rho v^{i})\,\)  as   a 4-vector field and the 4-momentum \(\pi_{\mu} = (\pi_{0},\pi_{i})\) as a covector field. In these terms the continuity equation has the form  \eq{01} and  the Euler equation \eq{1} appears  in a remarkably compact form. It follows
from \eq{028}  
\begin{align}
        \quad j^\mu\Omega_{\mu\nu}= 0\,,
 \la{pieq}
\end{align}
where \begin{align}
\Omega_{\mu\nu}=\p_{\mu}\pi_{\nu}-\p_{\nu}\pi_{\mu}
\end{align}
is the 4-vorticity  antisymmetric tensor  extended by electromagnetic field (also referred as canonical symplectic form or Khalatnikov canonical vorticity tensor see, e.g., \cite{G2,CarterKhalatnikov} and references therein). 

Euler equations in various forms (\ref{03},\ref{028},\ref{pieq}) are all equivalent. The advantage of the form \eq{pieq} is that it is insensitive to the space-time structure. For example, it stays the same regardless of whether the space-time is Galilean or Minkowski. The information
of the space-time structure is delegated to the relation between the fluid momentum \(\pi \) and the vector-current \(j\). For the Galilean barotropic fluid these relations are (\ref{pidef},{\ref{pi0def}). In the relativistic context, the Euler equation in the form \eq{pieq} is known as the Carter-Lichnerowitz equation (see \cite{G} for an excellent review of the subject). Eq.\eq{pieq} is the basis of geometric interpretation  of the Euler flow. It states that the vorticity vector field spans a bundle of integral two-dimensional surfaces normal to the space-time flow and that these surfaces form a  foliation of the space-time.

The spatial components of the 4-vorticity are: 
\begin{align}
        \Omega_{ij}=\e_{ijk} (\omega^k+B^k) \,.
\end{align} 
The Euler equation connects the space-time component \(\Omega_{0i}=\dot \pi_i-\p_i\pi_0\) to the spatial components by 
\be 
        \Omega_{0i}= v^j\Omega_{ji} \,.
 \la{24x}
\ee 
This relation states that 4-vorticity, similar to the density, is advected by the flow (the Helmholtz law): vorticity can not be destructed or created; instead, it moves with the flow. Equivalently, it implies that the Lie derivative of vorticity vanishes along the flow with the mass current $j^\mu$.

It is customary to present these equations in terms of differential forms. We assemble the 4-momentum 1-form \(\pi=\pi_\mu dx^\mu\) and  4-vorticity 2-form \(\Omega=\Omega_{\mu\nu} dx^\mu\wedge dx^\nu\), a  closed 2-form equal to the exterior derivative of the momentum
\begin{align}
        \Omega=d\pi\,,   \quad d\Omega=0 \,.
\end{align} 
The Euler equation in the form of Carter-Lichnerowitz \eq{pieq} is the statement that the 1-form obtained by the interior product between the current vector field and vorticity 2-form vanishes
\begin{align}
   \iota_{j}\Omega=0 \,.
 \la{ijdpi}
\end{align}

Now we turn to helicity. In the 4-dimensional formalism helicity is the 3-form
\begin{align}
        h=\pi\wedge d\pi=\pi\wedge\Omega \,.
 \la{28x}
\end{align}
The components of \(h\) are the helicity density \(h_0\) and the flux
\(\bm h\). They read
\begin{align}
        \begin{split}
        h_0 &=\bm\pi\!\cdot(\bm\nabla\!\times\!\bm\pi) \,,
 \\ 
        \bm h &= \bm \pi\times( \dot {\bm\pi}\!- \bm\nabla\pi_0)\!
        -\!\pi_0(\bm \nabla\!\times\!\bm\pi)
 \\&=h_0\bm v-(\bm\nabla\times \bm \pi)(\bm\pi \cdot\bm v+\pi_0)\,.
        \end{split}
 \la{20}
\end{align}
We comment that counter to vorticity, helicity is not frozen into the flow as \(\bm h\neq h_0\bm v\).  

We would like to show that the helicity 3-form \eq{28x} is closed 
\begin{align}
        dh=\Omega\wedge\Omega=0 \,.
 \la{26}
\end{align}
Eq.~\eq{26} amounts to the conservation of helicity \eq{06}  
\begin{align}
        \dot h_0+\bm\nabla\bm h=0 \,.
 \la{31x}
\end{align}
 One can check \eq{31x} either by elementary algebra with the help of the Euler equation or apply the following arguments. In four dimensions, $\Omega\wedge\Omega=d\pi \wedge d\pi$ is a 4-form, hence it is proportional to the 4-volume form. On the other hand, it follows from \eq{ijdpi} that $i_{j}(\Omega\wedge \Omega)=0$.  Assuming that the fluid density does not vanish and, therefore, $j\neq 0$, we conclude that the proportionality coefficient between $\Omega\wedge\Omega$ and the volume form is zero. This  implies the conservation of helicity in the form \eq{26}. 

To have the same argument in component notations we use the identity 
 \begin{align}
        &2\e^{\alpha\nu\lambda\rho} (\p_{\mu}\pi_{\nu}-\p_\nu\pi_\mu) \p_{\lambda}\pi_{\rho}
        =\delta^\alpha_\mu\e^{\tau\nu\lambda\rho}
        \p_{\tau}\pi_{\nu} \p_{\lambda}\pi_{\rho} \,.
 \la{i}
\end{align}
The RHS of the identity is \((\Omega\wedge\Omega)\delta^\alpha_\mu\).  
If we contract one  free index, say \(\mu\), of the LHS with the current \(j^{\mu}\), then the equation \eq{pieq} prompts that the contraction vanishes and we get \(0=j^\alpha (\Omega\wedge\Omega)\), hence \eq{26}.

 We comment that the derivation of the helicity conservation does not utilize the relation between the current $j$ and the momentum $\pi$. It  relies on  the Eq.\eq{pieq} (or \eq{ijdpi}) which states the  geometric property  of the flow:  surfaces spanned by the vorticity vector field orthogonal to the current form a foliation of the space time.
 
\paragraph{Axial anomaly.}   When we invoke the relation between the current and the momentum \eq{pidef}  we encounter a  caveat common to gauge theories. The canonical momentum $\pi$, and, therefore,
helicity 3-form  (\ref{28x},\ref{20}) are local in terms of the gauge potential but cannot be locally expressed through \(\bm E\) and \(\bm B\).  At the same time without electromagnetic field the 
helicity form \eq{28x}  is a local functional of the Eulerian fields 
\begin{align}
        \begin{split}
        &h_0=m\bm v\!\cdot\!\bm\omega \,,
        \quad \bm h=h_0\bm v
        -\bm\omega\left({m\bm v^2}+\Phi\right).
        \end{split}\la{35}
\end{align}[This form of helicity are  equivalent to the chirality \eq{20} up to exact form and identical to (\ref{9},\ref{8})].

It is desirable to extend the formulas \eq{35} for non-vanishing gauge potential in such a manner that they remain local in terms of electric and magnetic fields. The 4-dimensional formalism yields the result in few lines. In this framework the 4-chirality  is the 3-form 
\begin{align}
        j_A=(\pi-A)\wedge(d\pi+dA) \,,
 \la{34x}
\end{align}
where  \(A=A_\mu dx^\mu\) is the gauge potential 1-form. Components of the chirality form are defined by (\ref{9},\ref{8}). More precisely, \eq{34x}  yields the expression for the flux 
\be 
        \bm j_A=m\bm v \times (m\dot {\bm v}- \nabla\Phi-2\bm {E})-\Phi(\bm\omega+2\bm B)\,,
 \la{350}
\ee 
which under substitution \(m\dot{\bm v} \) from the Euler equation is identical to \eq{8}.

Helicity and chirality are related by the identity
\begin{align}
      \pi \wedge d\pi=j_A+A\wedge dA-d(A\wedge\pi) \,.
  \la{31}
\end{align}
We take the exterior derivative of both parts of \eq{31} and recall that the helicity is a closed form \eq{26}. We obtain the axial anomaly \eq{1} announced in the introduction in terms of  the chirality 3-form and the electromagnetic  2-form  \(F=dA=\tfrac 12F_{\mu\nu}dx^\mu
\wedge dx^\nu\) 
\be 
        dj_A=F\wedge F\,.
 \la{34}
\ee
  
This simple procedure explains the origin of the anomaly in the context of hydrodynamics: helicity \(\mathcal{H}\) is conserved, but its current \eq{28x} is not local in terms of Eulerian fields. At the same time chirality is comprised locally by \(\bm E\) and \(\bm B\) but  the Chern-Simons term \(A\wedge dA\)  in \eq{31} makes  its conservation  anomalous.  The exterior derivative of the Chern-Simons  is \(\bm E\cdot \bm B\). It  contributes the anomaly term in the divergence of the axial current.

\paragraph{Relativistic hydrodynamics.} The derivation above remains unchanged for \emph{relativistic} barotropic flow. Relativistic flow is characterized by the specific Gibbs energy  \(\mu_R = mc^2+\mu\) which differs from the internal chemical potential \(\mu\) by the rest energy, 4-velocity \(u_\mu\) normalized as \(u^\mu u_\mu=-1\), and canonical momentum 1-form \(\pi=(c^{-1}\mu_R u_\mu +A_\mu) dx^\mu\). In this case the mass current is \(j^\mu=n u^\mu\), where \(n\) is particle number. 

The formula for the chirality 3-form \eq{34x} remains intact and obeys the anomaly equation \eq{1}. In components, the chirality current reads
\begin{align}
    j_A^\mu=\e^{\mu\nu\lambda\rho}c^{-1}\mu_Ru_{\nu}\Big(c^{-1}\mu_{R}\p_\lambda u_\rho
    +  F_{\lambda\rho}\Big)
 \la{42} \,.
\end{align}
It is instructive to trace how does the compact relativistic expression \(\eq{42}\) turns into the cumbersome non-relativistic formula \eq{8}. The temporal component of \eq{42} \(j^0_A\) reproduces the chirality density (\ref{9}) as \(v/c\to 0\) in a straightforward manner. But the spatial component of \eq{42}, the flux, as it stands, does not have the non-relativistic limit. It diverges as \( j_A^i\to mc^{2}(\omega^i+2 B^i)\) with $c\to\infty$. However, the divergent term is a curl. We recall that the formula \eq{42} as well as its non-relativistic version \eq{350} is defined up to a curl. Hence, the divergent term can be dropped.  As a result, the non-relativistic limit of \eq{42} is a mix of terms originated from the rest energy in \(\mu_R\) and the expansion in \(v/c\). It yields  the expression which  differs from (15) by the curl $\bm\nabla\times (m\bm v \Phi)$ which does not affect the equation (14).

\paragraph{Summary.}  
Alike quantum field theories with Dirac fermions, Euler hydrodynamics, too, possesses a conserved vector current and an axial current, helicity. Both are conserved. The conservation of the vector current is explicitly imposed as the continuity equation and is associated with the gauge symmetry. The origin of helicity conservation is more subtle. It follows from the Euler equation and could be traced to different but ultimately related roots: a topological nature of helicity as a linkage of vortex lines, or to the degeneracy of the Poisson structure and its foliations, or to the {\it relabeling symmetry} acting in the extended phase space (see, e.g., \cite{Morrison, Zakharov} ). Our results indicate that these properties are represented by the axial gauge transformation \eq{002}. We defer these discussions pending further studies. 

These results also suggest that there may be special hydrodynamic flows representing
fermions of quantum field theory if the analogy is to be followed further.

From a hydrodynamics perspective, the origin of the anomaly may be summarized as follows. Electromagnetic forces do not destroy the conservation of helicity. However, the helicity current, although conserved (see, \eq{31x}), is expressed through the canonical fluid momentum. That prevents treating the helicity current as a local functional of \(\bm E\) and \(\bm B\). Similar to QED, the ``conflict'' is resolved at the expense of the conservation of the axial current. The chirality density defined by \eq{9}, is identical to the helicity density in the absence of electromagnetic fields. It is local in terms of \(\bm E\) and \(\bm B\) but is not conserved, obeying the equation \eq{09} identical to the axial anomaly equation \eq{1} of the quantum field theory. 

The appearance of the axial-current anomaly in 3D Euler flow is not accidental. All formulas presented in this work have a straightforward generalization to hydrodynamics in higher odd spatial dimensions matching the axial-current anomaly of quantum field theories with Dirac fermions. Also, we expect that not only perturbative anomalies but also global anomalies  appear in classical hydrodynamics. 

Emergence of the axial current anomaly in Euler fluids helps  to clarify and to illustrate aspects of anomalies in quantum field theories also  providing an interesting angle at fluid dynamics.}

\medskip

\noindent The authors thank  A. Cappelli and B. Khesin for helpful
discussions.  P.W. acknowledges helpful discussions with V.P. Nair, A.
Polyakov and G. Volovik. P.W. thanks G.  Volovik for pointing out the association
of the axial current anomaly with the fluid helicity discussed in \cite{V1} in the context of superfluid
$\operatorname{^3He-A}$. P.W. thanks Simons Center for Geometry and Physics
for the hospitality at various stages of this work. Both authors acknowledge
INFN\ Galileo Galilei Institute where the work had been completed. A.G.A.
thanks S. Vitouladitis for insightful comments.  The work of P.W. was supported
by the NSF under the Grant NSF DMR-1949963 and by the Ministry of Science and Higher Education of the Russian Federation, agreement No 075-15-2021-602. 
  The work of A.G.A. was supported by NSF DMR-1606591.


\end{document}